\documentclass[aps,twocolumn,prd,dvipsnames,usenames,floatfix,superscriptaddress,nofootinbib,longbibliography,preprintnumbers]{revtex4-1}

\usepackage{amsmath,amssymb,bm}

\usepackage{fancyhdr} 
\usepackage{footnote}
\makesavenoteenv{tabular}
\makesavenoteenv{table}
\usepackage[utf8x]{inputenc}
\usepackage{units}
\usepackage[english]{babel}
\usepackage{xspace}
\usepackage{paralist}
\usepackage{amsmath} 
\usepackage{float}
\usepackage{amssymb} 
\usepackage{graphicx}
\usepackage{bm}
\usepackage{tabularx}
\usepackage{slashed}
\usepackage[final,textsize=footnotesize,color=ForestGreen!05]{todonotes}

\newcommand{\eff}{\text{eff}}

\newcommand{\mn}{m_{N}}

\newcommand{\tn}{\tau_{N}}

\newcommand{\EM}{\text{EM}}
\newcommand{\neff}{\ensuremath{N_{\text{eff}}}\xspace}

\usepackage{subfig}
\usepackage[colorlinks=true,linktocpage=true,urlcolor=blue,citecolor=blue,linkcolor=blue]{hyperref}

\usepackage{dsfont}

\usepackage[justification=centerlast, format=plain, labelfont=bf,font=small]{caption}

\DeclareCaptionJustification{justified}{\leftskip=0pt \rightskip=0pt \parfillskip=0pt plus 1fil}

\makeatletter
\def\doauthor#1#2#3{%
  \ignorespaces#1\unskip
  \begingroup
   #3%
  \@if@empty{#2}{\@listcomma\endgroup{}{}}{\endgroup{\comma@space}{}\frontmatter@footnote{#2}}%
  \space \@listand
}%
\makeatother

\makeatletter
\def\@ssect@ltx#1#2#3#4#5#6[#7]#8{%
  \def\H@svsec{\phantomsection}%
  \@tempskipa #5\relax
  \@ifdim{\@tempskipa>\z@}{%
    \begingroup
      \interlinepenalty \@M
      #6{%
       \@ifundefined{@hangfroms@#1}{\@hang@froms}{\csname @hangfroms@#1\endcsname}%
       {\hskip#3\relax\H@svsec}{#8}%
      }%
      \@@par
    \endgroup
    \@ifundefined{#1smark}{\@gobble}{\csname #1smark\endcsname}{#7}%
  }{%
    \def\@svsechd{%
      #6{%
       \@ifundefined{@runin@tos@#1}{\@runin@tos}{\csname @runin@tos@#1\endcsname}%
       {\hskip#3\relax\H@svsec}{#8}%
      }%
      \@ifundefined{#1smark}{\@gobble}{\csname #1smark\endcsname}{#7}%
      \addcontentsline{toc}{#1}{\protect\numberline{}#8}%
    }%
  }%
  \@xsect{#5}%
}%
\makeatother

\makeatletter
\g@addto@macro\bfseries{\boldmath}
\makeatother

\begin{document}

\preprint{KCL-2021-05}
\author{Alexey Boyarsky$^{\mathds{A},}$}
\affiliation{Instituut-Lorentz for Theoretical Physics, Universiteit Leiden, Niels Bohrweg 2, 2333 CA Leiden, The Netherlands}
\author{Maksym Ovchynnikov$^{\mathds{M},}$}
\affiliation{Instituut-Lorentz for Theoretical Physics, Universiteit Leiden, Niels Bohrweg 2, 2333 CA Leiden, The Netherlands}
\author{Nashwan Sabti$^{\mathds{N},}$}
\affiliation{Department of Physics, King's College London, Strand, London WC2R 2LS, UK}
\author{Vsevolod Syvolap$^{\mathds{V},}$}
\affiliation{Niels Bohr Institute, University of Copenhagen,
Blegdamsvej 17, DK-2100 Copenhagen, Denmark}

\def\thefootnote{$\mathds{A}$\hspace{0.4pt}}\footnotetext{\href{mailto:boyarsky@lorentz.leidenuniv.nl}{boyarsky@lorentz.leidenuniv.nl}}
\def\thefootnote{$\mathds{M}$\hspace{0.pt}}\footnotetext{\href{mailto:ovchynnikov@lorentz.leidenuniv.nl}{ovchynnikov@lorentz.leidenuniv.nl}}
\def\thefootnote{$\mathds{N}$\hspace{0.7pt}}\footnotetext{\href{mailto:nashwan.sabti@kcl.ac.uk}{nashwan.sabti@kcl.ac.uk}}
\def\thefootnote{$\mathds{V}$\hspace{0.7pt}}\footnotetext{\href{mailto:vsevolod.syvolap@nbi.ku.dk}{vsevolod.syvolap@nbi.ku.dk}}
\setcounter{footnote}{0}
\def\thefootnote{\arabic{footnote}}

\title{When FIMPs Decay into Neutrinos: The $N_\mathrm{eff}$ Story}

\begin{abstract}
\noindent The existence of feebly interacting massive particles (FIMPs) could have significant implications on the effective number of relativistic species $N_\mathrm{eff}$ in the early Universe. In this work, we investigate in detail how short-lived FIMPs that can decay into neutrinos affect  $N_\mathrm{eff}$ and highlight the relevant effects that govern its evolution. We show that even if unstable FIMPs inject most of their energy into neutrinos, they may still decrease $N_{\mathrm{eff}}$, and identify neutrino spectral distortions as the driving power behind this effect. As a case study, we consider Heavy Neutral Leptons (HNLs) and indicate which regions of their parameter space increase or decrease $N_{\mathrm{eff}}$. Moreover, we derive bounds on the HNL lifetime from the Cosmic Microwave Background and comment on the possible role that HNLs could play in alleviating the Hubble tension.
\end{abstract}

\maketitle

\section{Introduction}
\label{sec:introduction}

The Standard Model (SM) of particle physics and cosmology has proven to be very successful in describing our observable Universe. Nevertheless, in its current form, it does not provide a physical origin for a number of observed phenomena. The baryon asymmetry of the Universe, the existence of neutrino masses and the evidence for the dark matter each establish that the SM picture is not complete and call for the addition of new physics, usually in the form of a new particle~\cite{Canetti:2012zc, Strumia:2006db, Bertone:2004pz}. While the landscape of viable candidates is almost limitless, extensions of the SM often involve feebly interacting massive particles (FIMPs) that possess small couplings to the SM sector, see e.g.~\cite{Baer:2014eja,Bernal:2017kxu, Lanfranchi:2020crw, Agrawal:2021dbo} for reviews. Their inclusion in the primordial plasma comes with potential consequences on fundamental probes in the early Universe, such as primordial nucleosynthesis and the Cosmic Microwave Background (CMB)~\cite{Sarkar:1995dd, Chen:2003gz, Pospelov:2010hj}. Indeed, just their presence in the system would already contribute to the dynamics of the expanding Universe, let alone further implications that could follow from their decay.

\vspace{10pt}

A key parameter in this topic is the effective number of relativistic species \neff, given by:
\begin{align}
    \label{eq:Neff}
    N_{\rm eff} \equiv \frac{8}{7}\left(\frac{11}{4} \right)^{4/3} \left( \frac{\rho_\text{rad}-\rho_\gamma}{\rho_\gamma}\right)\ ,
\end{align}
where $\rho_\mathrm{rad}$ and $\rho_\gamma$ are the total radiation and photon energy densities respectively. We define the change in this quantity as $\Delta\neff = \neff - \neff^\mathrm{SM}$, where within the Standard Model $\neff^\mathrm{SM} \simeq 3.044$~\cite{Bennett:2019ewm, Escudero:2020dfa, Bennett:2020zkv, Froustey:2020mcq, Akita:2020szl}. Any deviation from the SM value is regulated by weak interactions between {\parfillskip=0pt\par}

neutrinos and electromagnetic (EM) particles, which are efficient enough at temperatures $T\gg1\,\mathrm{MeV}$ to keep these species in equilibrium with each other. At lower temperatures, the interactions gradually go out of equilibrium and the energy exchange between the two sectors will stop. Decaying FIMPs can affect this delicate process in different ways, depending on whether they inject most of their energy into EM particles or neutrinos. 

\vspace{10pt}

The impact of FIMPs predominantly decaying into EM particles has been extensively studied in the literature, see e.g.~\cite{Kawasaki:1999na,Kawasaki:2000en,Kawasaki:2007mk,Hasegawa:2019jsa,Fradette:2018hhl,Sabti:2020yrt}. Such particles heat up the EM plasma and consequently decrease \neff, independently of whether the decay happens during or after neutrino decoupling. 
On the other hand, for FIMPs that mostly decay into neutrinos, we naively expect that \neff would increase. This is indeed true for lifetimes $\tau_\mathrm{FIMP} \gg t_{\nu}^\mathrm{dec} \sim 0.1-1\,\mathrm{s}$, where $t_{\nu}^\mathrm{dec}$ is the time of neutrino decoupling, see e.g.~\cite{Fuller:2011qy}. However, the situation for lifetimes $\tau_\mathrm{FIMP} \sim t_{\nu}^\mathrm{dec}$ is different: the neutrinos are still in partial equilibrium and try to equilibrate with the injected neutrinos. This scenario has been considered before in~\cite{Hannestad:2004px,Dolgov:2000jw,Ruchayskiy:2012si} that arrived at different conclusions about the impact on \neff. The work~\cite{Hannestad:2004px} studied a reheating scenario in which all the SM particles are absent before FIMPs start decaying. In such a framework, all neutrinos have high energies, which means that they mainly thermalise via neutrino-EM interactions and \neff naturally decreases. References~\cite{Dolgov:2000jw,Ruchayskiy:2012si} considered Heavy Neutral Leptons (HNLs) with masses $\mn < m_{\pi}$ and lifetimes $\tn \lesssim 1\, \text{s}$. Such HNLs are in thermal equilibrium in the early Universe, but decouple as the Universe expands and eventually decay mainly into high-energy neutrinos at MeV temperatures. These two works drew different conclusions about \neff: \cite{Dolgov:2000jw} reported $\Delta \neff > 0$ for the whole studied mass range, whereas~\cite{Ruchayskiy:2012si} presented in their Fig.~3 that $\Delta \neff < 0$ for masses $60\text{ MeV}\lesssim\mn <m_{\pi}$ and lifetimes $\tau_N \ll 1\, \mathrm{s}$. The sign of $\Delta \neff $ is not emphasised in these two papers;~\cite{Ruchayskiy:2012si} did not comment on the contradiction with~\cite{Dolgov:2000jw} on this issue and no physical discussion of this phenomenon was provided\footnote{A more recent work~\cite{Gelmini:2020ekg} considered \emph{long-lived} (i.e., decaying after $e^{+}e^{-}$ annihilation) HNLs that could decay both into EM particles and neutrinos. In this case, \neff could both increase and decrease, as at such late times the injected energy densities from HNL decays can dominate over the SM densities of both the EM and neutrino sectors. Another recent paper~\cite{Mastrototaro:2021wzl}, which appeared after our work was submitted, claims that $\Delta \neff \geq 0$ for all cases in which FIMPs decay mostly into neutrinos. We comment on it in Appendix~\ref{app:previous_literature}.}.

\vspace{10pt}

In this paper, we aim to clarify the behaviour of \neff in the presence of FIMPs that decay mainly into neutrinos and have lifetimes $\tau_\mathrm{FIMP} \sim t_{\nu}^\mathrm{dec}$. Here we will assume that a thermal bath of SM particles is already present in the primordial Universe. We will first construct a simple model in Sec.~\ref{sec:qualitative-analysis} that provides us with a qualitative understanding of how such particles impact \neff, the findings of which we then confirm by using the Boltzmann code \texttt{pyBBN}~\cite{Sabti:2020yrt}. Our analysis shows that short-lived FIMPs that inject most of their energy into neutrinos may decrease \neff. This is because during the equilibration process, the injected high-energy neutrinos redistribute their energy among the neutrino and EM plasma. If the energy of the injected neutrinos is sufficiently large, the energy transfer to the EM sector occurs faster than the equilibration with the neutrino sector. This means that the EM plasma heats up more than the neutrino plasma, which eventually leads to $\Delta\neff < 0$. We will find that this mechanism is especially relevant for FIMPs with masses larger than a few tens of MeV. We will then apply these general considerations to the well-motivated case of Heavy Neutral Leptons. This is done in Sec.~\ref{sec:HNLs}, where we will also briefly discuss their influence on the physics at the CMB epoch, derive bounds on their mass and lifetime, and comment on the implications their presence may have for the Hubble tension. Our main results are summarised in Figs.~\ref{fig:Neff_analytic} and~\ref{fig:HNL_parameterspace}. Finally, we will present our conclusions in Sec.~\ref{sec:conclusions}. Complementary details and simulation results are included in the appendices~\ref{app:residual_spectral}$-$\ref{app:HNL_fitting_functions}.

\section{How FIMPs impact \neff}
\label{sec:qualitative-analysis}
The goal of this section is to provide a physical understanding of the processes that govern the behaviour of \neff in the presence of decaying FIMPs. In particular, we focus on FIMPs with masses $\gg 1\,\mathrm{MeV}$ that decay when neutrinos are still in (partial) equilibrium. Such FIMPs can decay into high-energy neutrinos that then participate in interactions with thermal neutrinos and electrons/positrons. We will find that even if most of the FIMP energy is injected into neutrinos, these interactions may still cause a decrease in \neff. This feature appears since the injected high-energy neutrinos get quickly converted into electrons/positrons and drag thermal neutrinos residing in the plasma along with them. During this process, neutrino-neutrino interactions lead to the presence of residual non-thermal distortions in the distribution functions of neutrinos (neutrino spectral distortions) that keep the balance of $\nu\leftrightarrow\mathrm{EM}$ interactions shifted to the right till long after the injection (i.e., more energy is transferred from the neutrino plasma to the electromagnetic plasma than vice versa). The energy transfer from neutrinos to EM particles accumulated over time can then be sizeable enough, such that $\Delta\neff$ becomes negative. This effect diminishes with larger FIMP lifetime, as neutrino-EM interactions go out of equilibrium and neutrinos can no longer be converted into electrons/positrons. Therefore, FIMPs that decay into neutrinos after neutrino decoupling will lead to $\Delta\neff > 0$. In what follows, we will consider FIMPs that can decay into both neutrinos and EM particles, and construct a simple model that provides a semi-analytic description of the aforementioned effect. At the end of this section and in Appendix~\ref{app:residual_spectral}, we will also highlight and further elaborate on the central role of neutrino spectral distortions in the dynamics of \neff.

\vspace{10pt}

A FIMP of mass $m_{\text{FIMP}}\gg T$ can decay into neutrinos with energies that are larger than those found in the primordial plasma. Such non-equilibrium neutrinos manifest themselves as spectral distortions at high energies and will subsequently interact with the thermal neutrinos and electrons/positrons in the plasma. We assume that the amount of injected non-equilibrium neutrinos is only a small fraction of the thermal neutrinos in the plasma. The evolution of the injected neutrinos is then mainly governed by the following reactions:
\begin{align}
    & \nu_{\text{non-eq}} + \nu_{\text{therm}} \to \nu_{\text{non-eq}} + \nu_{\text{non-eq}}
    \label{eq:therm-proc-1} \\
    & \nu_{\text{non-eq}} + \overline{\nu}_{\text{therm}} \to e^{+} + e^{-}
        \label{eq:therm-proc-2} \\
    & \nu_{\text{non-eq}} + e^{\pm} \to \nu_{\text{non-eq}} + e^{\pm}\ ,
        \label{eq:therm-proc-3}
\end{align}
where `non-eq' and `therm' refer to neutrinos with non-equilibrium and thermal energies respectively. Through these reactions, non-equilibrium neutrinos thermalise and quickly redistribute their energy among the neutrino and EM plasma. The energy loss rate of these non-equilibrium neutrinos is higher than the interaction rates of thermal particles~\cite{Dolgov:2002wy}:
\begin{align}
\label{eq:noneq_eq_rates}
\frac{\Gamma_{\text{non-eq}}}{\Gamma_{\text{therm}}} \sim \frac{G_{F}^{2}T^{4}E_{\nu}^\mathrm{inj}}{G_{F}^{2}T^{5}} = \frac{E_{\nu}^\mathrm{inj}}{T} \gg 1\ ,
\end{align}
where $E_\nu^\mathrm{inj}$ is the average energy of the injected non-equilibrium neutrinos. Note that reactions between thermal particles also exchange energy between the neutrino and EM sectors, but this energy exchange is subdominant as far as Eq.~\eqref{eq:noneq_eq_rates} holds.

\vspace{10pt}

The amount of energy that ends up in the EM plasma has three contributions: \emph{1)} the direct decay of FIMPs into EM particles, \emph{2)} the energy transfer of non-equilibrium neutrinos to EM particles during thermalisation and \emph{3)} the energy transfer from thermal neutrinos to EM particles as a consequence of them being dragged by non-equilibrium neutrinos during thermalisation (reactions~\eqref{eq:therm-proc-1} and~\eqref{eq:therm-proc-2}). The first process injects a fraction $\xi_\EM$ of the total FIMP energy into the EM plasma, while the latter two increase this fraction to:
\begin{align}
\label{eq:fEM_eff}
\xi_{\EM,\eff}(E_\nu^\mathrm{inj},T) = \xi_{\EM} + \xi_\nu\times \epsilon(E_\nu^\mathrm{inj}, T)\ ,
\end{align}

where $\xi_\nu = 1 - \xi_{\EM}$ is the energy fraction that FIMPs directly inject into the neutrino sector and $\epsilon = \epsilon_\text{non-eq} + \epsilon_\text{thermal}$ is the effective fraction of $\xi_\nu$ that went to the EM plasma during the thermalisation. The latter quantity can be split in a contribution from non-equilibrium neutrinos ($\epsilon_\text{non-eq} = E_\nu^{\text{non-eq}\to\mathrm{EM}}/E_\nu^\text{inj}$) and an \emph{effective} contribution from thermal neutrinos ($\epsilon_\text{thermal} = E_\nu^{\text{thermal}\to\mathrm{EM}}/E_\nu^\text{inj}$).

\vspace{10pt}

Now, based on Eq.~\eqref{eq:fEM_eff}, if $\epsilon > 0.5$, then $\xi_{\EM,\eff} > 0.5$. This means that more than half of the FIMP energy eventually ends up in the EM plasma (i.e., EM plasma heats up more than the neutrino plasma), which results in $\Delta \neff < 0$ independently of the value of $\xi_{\EM}$\footnote{Note that it is not a requirement that $\epsilon$ must be larger than 0.5 in order for $\Delta \neff$ to be negative. It only signifies the independence from $\xi_\mathrm{EM}$.}. This simplified energy redistribution picture only holds if the non-equilibrium neutrino energy is much larger than the average energy of thermal neutrinos. Once these two energies become similar in magnitude, backreactions cannot be neglected anymore and the evolution can only be accurately described with a system of Boltzmann equations.

\vspace{10pt}

We can make a simple estimate of $\epsilon$ as a function of the injected neutrino energy $E_\nu^\mathrm{inj}$ and temperature $T$. We start with describing the thermalisation process of a \emph{single} injected neutrino, which causes a cascade of non-equilibrium neutrinos. Such a cascade can result after the injected neutrino participates in the processes~\eqref{eq:therm-proc-1}$-$\eqref{eq:therm-proc-3}. We assume that in the processes~\eqref{eq:therm-proc-1} and~\eqref{eq:therm-proc-3} each non-equilibrium neutrino in the final state carries half of the energy of the non-equilibrium neutrino in the initial state. Thus, roughly speaking, the thermalisation occurs during $N_{\text{therm}} \simeq \log_{2}(E_\nu^\text{inj}/3.15T)$ interactions. In addition, the process~\eqref{eq:therm-proc-1} doubles the number of non-equilibrium neutrinos, while~\eqref{eq:therm-proc-2} makes neutrinos disappear and~\eqref{eq:therm-proc-3} leaves the number unchanged. Therefore, after the $k$-th step in the cascade, the average number of non-equilibrium neutrinos is given by:

\begin{align}
        N_{\nu}^{(k)} &= N_{\nu}^{(k-1)}\left(2P_{\nu\nu\to\nu\nu} + P_{\nu e\to\nu e}\right) \nonumber\\
        &= N_\nu^{(0)}\left(2P_{\nu\nu\to\nu\nu} + P_{\nu e\to\nu e}\right)^{k}\ ,
\end{align}

with $N_\nu^{(0)} = 1$, and the total non-equilibrium energy is:
\begin{align}
       E_{\nu}^{(k)} &= E_{\nu}^{(k-1)}\left(P_{\nu\nu\to\nu\nu} +\frac{1}{2}P_{\nu e\to\nu e}\right) \nonumber\\ 
       &= E_\nu^\text{inj}\left(P_{\nu\nu\to\nu\nu} +\frac{1}{2}P_{\nu e\to\nu e}\right)^{k}\ ,
       \label{eq:non-eq-energy-cascade}
\end{align}

where $P_{\nu\nu\to\nu\nu},\, P_{\nu\nu\to ee},\,\text{and}\, P_{\nu e\to\nu e}$ are the average probabilities of the processes~\eqref{eq:therm-proc-1}$-$\eqref{eq:therm-proc-3}, respectively, and their sum equals unity. We define these probabilities as $P_i = \Gamma_i/\Gamma_\nu^\mathrm{tot}$, where $\Gamma_i$ is the interaction rate of each process and $\Gamma_\nu^\mathrm{tot}$ is the total neutrino interaction rate. The relevant reactions and their corresponding matrix elements are summarised in appendix D of~\cite{Sabti:2020yrt}. Assuming a Fermi-Dirac distribution for neutrinos and averaging over neutrino flavours, we find: 
\begin{align}
\label{eq:probabilities}
        P_{\nu\nu\to\nu\nu} \approx 0.76,\quad P_{\nu\nu\to ee} \approx 0.05,\quad P_{\nu e\to\nu e} \approx 0.19\, .
\end{align}

Finally, the value of $\epsilon_\text{non-eq}$ that accounts for the energy transfer from non-equilibrium neutrinos to the EM plasma is given by:
\begin{align}
        \epsilon_{\text{non-eq}} & = \frac{1}{E_\nu^\text{inj}}\sum_{k = 0}^{N_{\text{therm}}}\left(\frac{P_{\nu e\to\nu e}}{2} + P_{\nu\nu\to ee}\right)E_{\nu}^{(k)}\ .
        \label{eq:epsilon-noneq}
\end{align} 

In addition to the transferred non-equilibrium energy, the non-equilibrium neutrinos catalyse the energy transfer from thermal neutrinos to the EM plasma via the processes~\eqref{eq:therm-proc-1} and~\eqref{eq:therm-proc-2}. In other words, during the thermalisation process non-equilibrium neutrinos drag thermal neutrinos along with them, which leads to part of the energy stored in the thermal neutrino sector to end up in the EM sector. We assume that each reaction~\eqref{eq:therm-proc-1} transfers an energy amount of $3.15T$ from the thermal neutrino sector to non-equilibrium neutrinos, which then via~\eqref{eq:therm-proc-2} ends up in the EM plasma. Moreover, each reaction~\eqref{eq:therm-proc-2} contributes to another energy transfer of $3.15T$ from thermal neutrinos to the EM plasma. The effective contribution coming from this transfer is therefore:
\begin{align}
    \epsilon_{\text{thermal}} =\ & 
    \frac{3.15T}{E_\nu^\text{inj}}N_\nu^{\text{therm}\to\mathrm{EM}}\nonumber\\
    =\ & \frac{3.15T}{E_\nu^\text{inj}}P_{\nu\nu\to ee}\Bigg(\sum_{k = 0}^{N_{\text{therm}}}N_{\nu}^{(k)}\nonumber\\ 
    & + \left[P_{\nu\nu\to\nu\nu} + \sum_{k = 1}^{N_{\text{therm}}}\left(2P_{\nu\nu\to\nu\nu}\right)^{(k)}\right]\Bigg)\ ,
    \label{eq:epsilon-thermal}
\end{align}

where the first term in the round brackets is the contribution from the process~\eqref{eq:therm-proc-2} and the terms in the square brackets are the contribution from the process~\eqref{eq:therm-proc-1}. Note that the factor of 2 in the second sum accounts for the doubling of non-equilibrium neutrinos in the process~\eqref{eq:therm-proc-1}. We find that $\epsilon_{\text{thermal}}$ is at least 5 times smaller than $\epsilon_{\text{non-eq}}$, which makes this a sub-dominant effect.

\vspace{10pt}

\begin{figure}[t!]
    \vspace{-2.5pt}
    \centering
    \includegraphics[width=\linewidth]{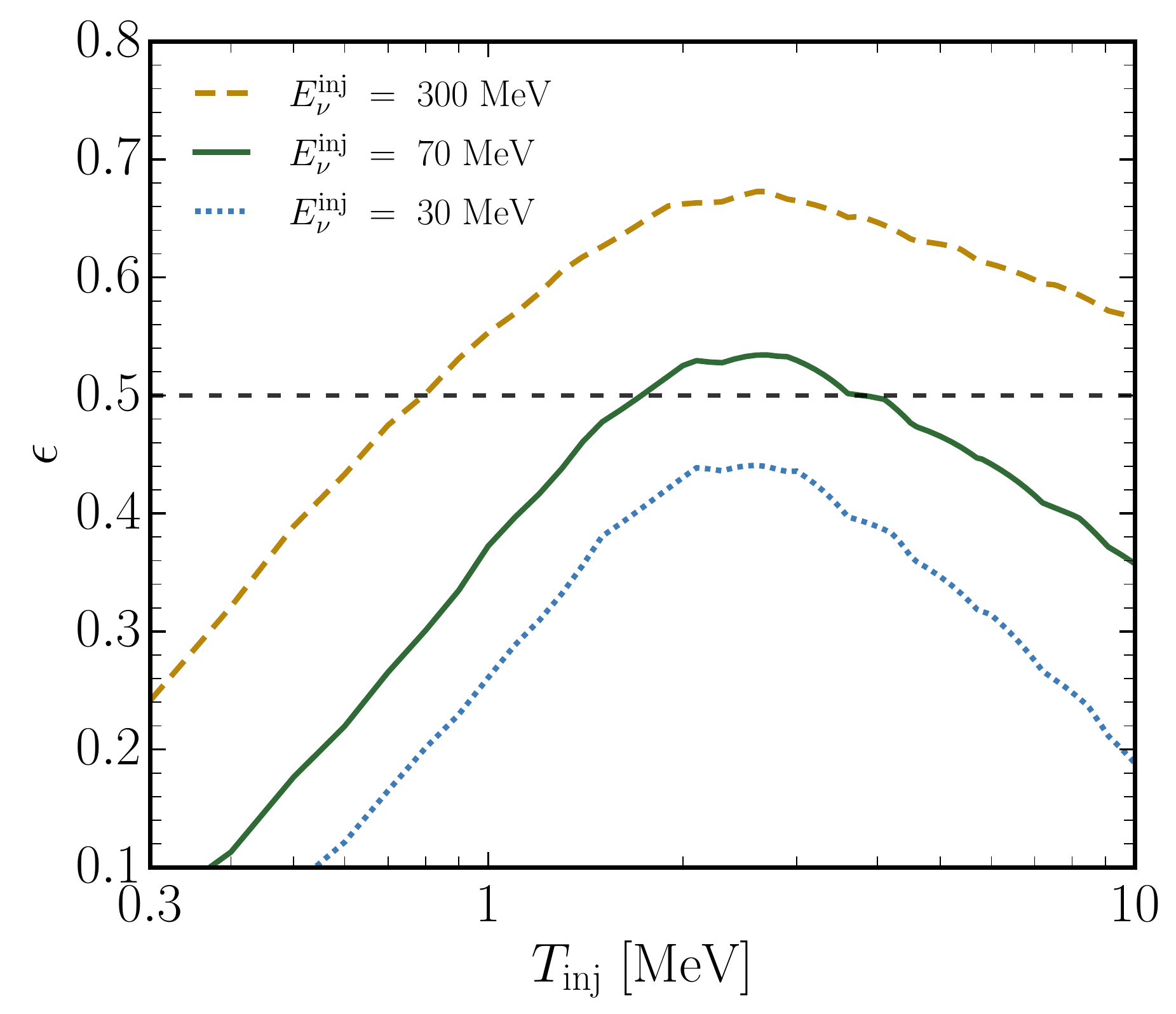}
    \caption{Estimate of the fraction of injected neutrino energy $\epsilon$ (both thermal and non-equilibrium) that gets transferred to the EM plasma during thermalisation (see text for details). The three curves indicate the value of $\epsilon$ when a neutrino of energy $E_\nu^\mathrm{inj}$ is injected at a temperature $T_\mathrm{inj}$. At high temperatures of order of $T_\mathrm{inj}\simeq E_\nu^\text{inj}$, the injected neutrinos are thermal-like, and hence $\epsilon$ is small. Once the temperature decreases, we enter the regime $E_\nu^\text{inj} \gg 3.15T_\mathrm{inj}$ and neutrinos transfer a significant amount of their energy to the EM plasma while thermalising. With further decrease of $T_\mathrm{inj}$, weak reactions go out of equilibrium and the energy transfer becomes less and less efficient, which results in a quick drop-off of $\epsilon$.}
    \label{fig:epsilon-behavior}
\end{figure}

As the Universe expands and the temperature decreases, weak reaction rates start to compete with the Hubble rate $H$. The energy transfer from neutrinos to the EM plasma therefore becomes less and less efficient, and $\epsilon$ tends to zero. In order to incorporate this effect, we multiply the probabilities in~\eqref{eq:probabilities} with a factor $\text{min}[\Gamma_i/H, 1]$, where $\Gamma_i = \Gamma_i(E_\nu^\text{inj}/2^{k})$ is the interaction rate of any of the processes~\eqref{eq:therm-proc-1}$-$\eqref{eq:therm-proc-3}. The resulting energy fraction of neutrinos that is transferred to the EM plasma $\epsilon = \epsilon_{\text{non-eq}}+\epsilon_{\text{thermal}}$ is shown in Fig.~\ref{fig:epsilon-behavior} for a number of injected neutrino energies $E_\nu^\text{inj}$. We find that $\epsilon$ can exceed 0.5 for $E_\nu^\mathrm{inj} \gtrsim 60\,\mathrm{MeV}$. This means that when FIMPs decay into neutrinos with such energies at temperatures of a few MeV, the majority of the injected neutrino energy will end up in the EM plasma during the thermalisation. This then leads to a decrease of \neff, independently of how much energy the FIMPs inject into the EM sector. 

\vspace{10pt}

\begin{figure}[t!]
    \vspace{-5pt}
    \centering
    \includegraphics[width=\linewidth]{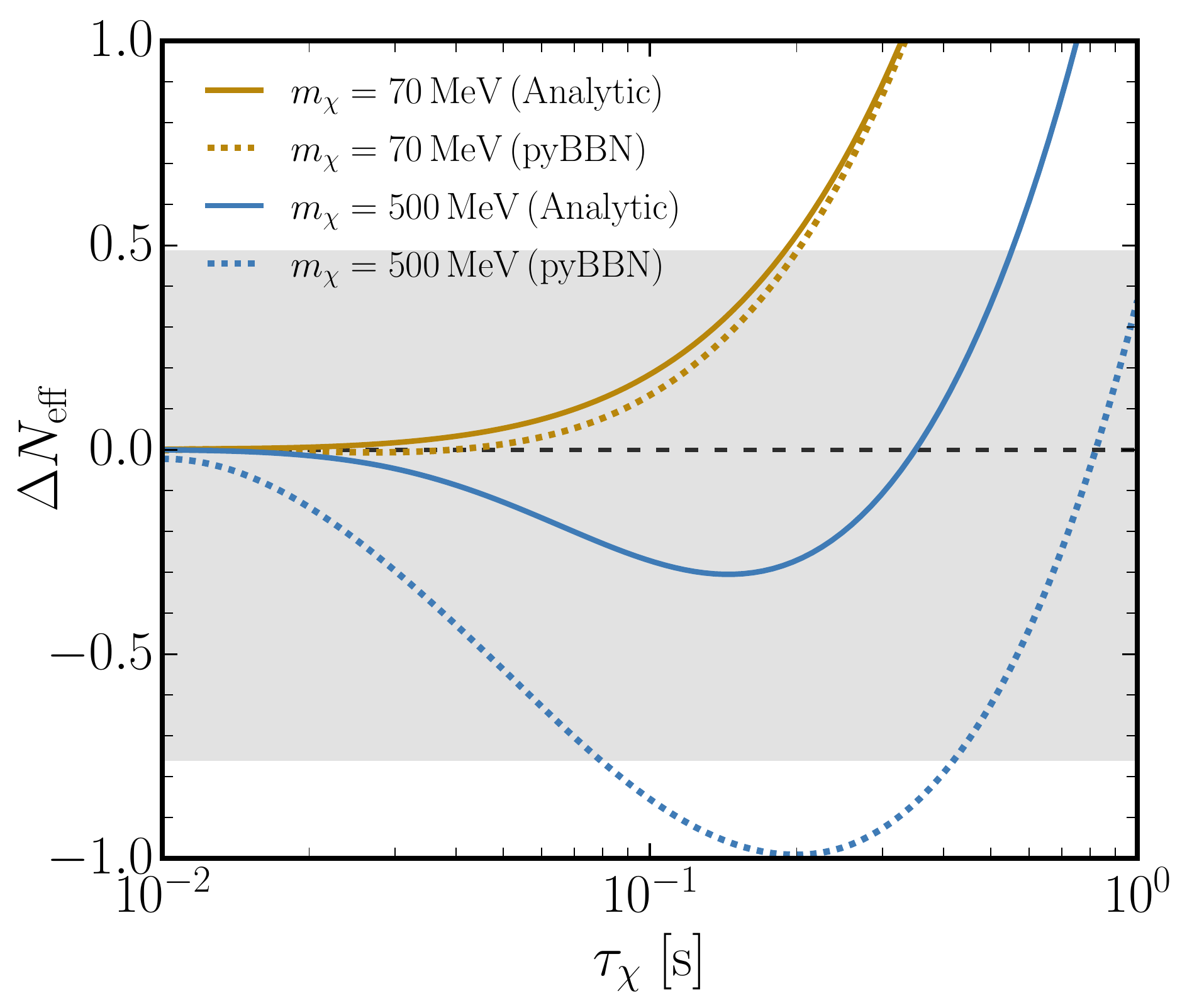}
    \caption{$\Delta\neff$ as a function of the lifetime of a FIMP $\chi$ that can only decay into neutrinos through $\chi \rightarrow \nu_e + \nu_\mu + \overline{\nu}_\mu$. The initial FIMP abundance is assumed to be $n_\chi/s = 0.01$ at $T = 1\,\mathrm{GeV}$, where $s$ is the total entropy density of a universe consisting of photons, neutrinos and electrons/positrons. The solid lines are the result of our semi-analytic model, while the dotted lines are obtained with the Boltzmann code \texttt{pyBBN}. The grey band is the current sensitivity by Planck (see text for details). The golden curves roughly indicate the lowest FIMP mass for which \neff can decrease due to the thermalization of the injected neutrinos. The stronger decrease of the blue, dotted curve as compared to the solid curve highlights the significance of residual neutrino spectral distortions in the evolution of $N_\mathrm{eff}$ (see Appendix~\ref{app:residual_spectral} for more details).}
    \label{fig:Neff_analytic}
\end{figure}

Now that we are able to estimate $\epsilon$, we can compute the correction to \neff for some benchmark FIMP scenario. It is worth noting here again that $\epsilon$ only depends on the energy of the injected neutrino and the temperature at which the injection happens. This means that $\epsilon$ is an independent quantity of the FIMP model considered, in contrast to $\xi_\mathrm{EM}$ and $\xi_\nu$, which do depend on the choice of the model. As an illustrative example, we assume that $\xi_\mathrm{EM} = 0$, i.e., the FIMP injects all of its energy into neutrinos ($\xi_\nu = 1$). Given that in our simple model neutrinos thermalise very quickly, we assume that they have a thermal-like distribution with a temperature $T_\nu$ and follow the approach in~\cite{Escudero:2018mvt,Escudero:2020dfa} to obtain the time evolution of $T_\nu$ and $T_\mathrm{EM}$ in the presence of decaying FIMPs (see Appendix~\ref{app:T_eqs}, where we provide the relevant equations). In this benchmark example, we consider a generic FIMP of mass $500\,\mathrm{MeV}$ that can decay only into three neutrinos and show $\Delta\neff$ as a function of its lifetime in Fig.~\ref{fig:Neff_analytic}. In order to compare the accuracy of our simple model, we also include in this figure the evolution of $\Delta\neff$ as obtained from the publicly available Boltzmann code \texttt{pyBBN}\footnote{\href{https://github.com/ckald/pyBBN}{https://github.com/ckald/pyBBN}}~\cite{Sabti:2020yrt}. The grey band in this figure indicates the current sensitivity of $N_\mathrm{eff}$ by Planck, which at 2$\sigma$ reads\footnote{This value is obtained from the Planck 2018 baseline TTTEEE+lowE analysis, where \neff, $Y_\mathrm{P}$ and the six base parameters in $\Lambda$CDM are varied.} $N_\mathrm{eff}^\mathrm{CMB} = 2.89 \pm 0.62$~\cite{Aghanim:2018eyx, Aghanim:2019ame}. We see that \neff can significantly decrease as a result of the thermalisation of the injected neutrinos. This decrease of \neff would only be further amplified if the FIMPs were also to inject some of their energy into the EM plasma. Importantly, we find that when using the Boltzmann equation $N_\mathrm{eff}$ decreases more than predicted by our semi-analytic model. This is because our model assumes that the remaining fraction $1-\epsilon$ of the injected neutrinos is perfectly thermal. In fact, this is usually not the case and the remaining non-equilibrium neutrinos manifest themselves as residual spectral distortions in the distribution function of neutrinos that further lead to a transfer of energy from the neutrino sector to the EM sector. We see that in some cases the inclusion of this effect can make the difference between being excluded by current data or not. We elaborate more on the effect of spectral distortions in Appendix~\ref{app:residual_spectral}. In short, the semi-analytic model is useful in providing a qualitative understanding of the behaviour of \neff in the presence of decaying FIMPs. On the other hand, if the aim is to obtain accurate predictions for \neff (relevant for setting bounds and forecasting), it is crucial to use the Boltzmann equation to track the evolution of the neutrino distribution functions. As such, we will use \texttt{pyBBN} in the remainder of this paper to simulate the impact of FIMPs on \neff.

\vspace{10pt}

As a final point, we can make a rough model-independent estimate for which neutrino energies the decrease of \neff happens. In the particular FIMP scenario considered here, we find that this effect occurs for masses higher than ${\sim}70\,\mathrm{MeV}$ (see Fig.~\ref{fig:Neff_analytic}). Given that in this case the neutrinos are created via 3-body decays, this would correspond to an average injected neutrino energy of roughly $E_\nu^\mathrm{inj}\sim m_\mathrm{FIMP}/3 \sim 25\,\mathrm{MeV}$. Therefore, as long as a FIMP injects most of its energy into neutrinos around neutrino decoupling, \neff could decrease if neutrinos with energies of at least $E_\nu^\mathrm{inj}\sim 25\,\mathrm{MeV}$ are produced.

\section{Case Study: Heavy Neutral Leptons}
\label{sec:HNLs}

A specific class of FIMPs that has seen an increasing interest over the last few years involves Heavy Neutral Leptons (HNLs). HNLs can be regarded as the right-handed companions to the SM neutrinos, which in a natural way give rise to neutrino masses~\cite{Ma:1998dn, Mohapatra:2006gs}. Besides this, they could play the role of the dark matter~\cite{Adhikari:2016bei,Boyarsky:2018tvu} and provide a mechanism for the generation of the baryon asymmetry in the Universe~\cite{Akhmedov:1998qx, Canetti:2012kh, Asaka:2017rdj}.
We consider HNLs that couple to the SM through the neutrino portal:
\begin{align}
  \label{eq:neutrino_portal}
  \mathcal{L}_\text{neutrino portal} = F_{\alpha\beta}(\overline{L}_\alpha\widetilde{H})N_\beta  + \mathrm{h.c.}\, ,
\end{align}

where $\alpha = \{e, \mu, \tau\}$, $\beta = 1,2,3,\ldots$, $F_{\alpha\beta}$ are dimen\--{\parfillskip=0pt\par}

sionless Yukawa couplings, $L_\alpha$ is the SM lepton doublet, $\widetilde{H} \equiv i\sigma_{2}H^{*}$ is the Higgs doublet in the conjugated representation and $N_\beta$ are the HNLs. This portal induces an effective interaction with SM particles that is similar to that of active neutrinos, but with an additional suppression in the form of small mixing angles~\cite{Bondarenko:2018ptm}. The mixing angle also parametrises with which neutrino flavour generation HNLs mix. An extension of the SM with three HNLs is also known as the neutrino Minimal Standard Model ($\nu$MSM)~\cite{Asaka:2005an, Asaka:2005pn, Shaposhnikov:2008pf}. This framework includes one light HNL, that plays the role of the dark matter, and two heavier ones, which are able to account for neutrino masses and the baryon asymmetry.

\vspace{10pt}

In what follows, we will consider two quasi-degenerate HNLs~\cite{Shaposhnikov:2006nn, Gavela:2009cd}, with masses up to $m_N\sim 1\,\mathrm{GeV}$ and lifetimes down to $0.01\,\mathrm{s}$. Higher masses are not considered, as we do not include decays into multi-meson final states that become relevant in this mass range~\cite{Bondarenko:2018ptm} and for which currently there is no adequate description of the corresponding decay widths\footnote{This makes it complicated to compute $\xi_{\nu}$ (and thus $\Delta \neff$), as it depends on the branching ratios of the different multi-meson decay channels. For instance, the decay $N\to 3\pi^{0}+\nu$ injects more energy into the EM plasma and diminishes $\xi_\nu$, while $N\to 3\pi^{\pm}+\ell^\mp$ may inject more energy into neutrinos and compensate for this decrease. Therefore, both such channels should be accounted for.}. \hspace{-1pt}Moreover, HNLs with lifetimes shorter than $0.01\,\mathrm{s}$ change \neff well below our accuracy (which is at the 1\% level). We make use of \texttt{pyBBN}~\cite{Sabti:2020yrt} to simulate their impact on the cosmological history, in particular on \neff and the primordial helium abundance $Y_\mathrm{P}$. We examine the region of parameter space in which HNLs inject most of their energy into neutrinos, but where $\Delta\neff$ is negative, illustrating the effect described in the previous section. Finally, we derive bounds from the CMB and comment on the possible role of HNLs in alleviating the Hubble tension. In Appendix~\ref{app:HNL_fitting_functions}, we include fitting functions for \neff as obtained from the simulations.


\subsection{Impact on Cosmological History}
\label{subsec:HNLs_Neff}

\begin{figure}[t!]
    \vspace{-2pt}
    \centering
    \includegraphics[width=\linewidth]{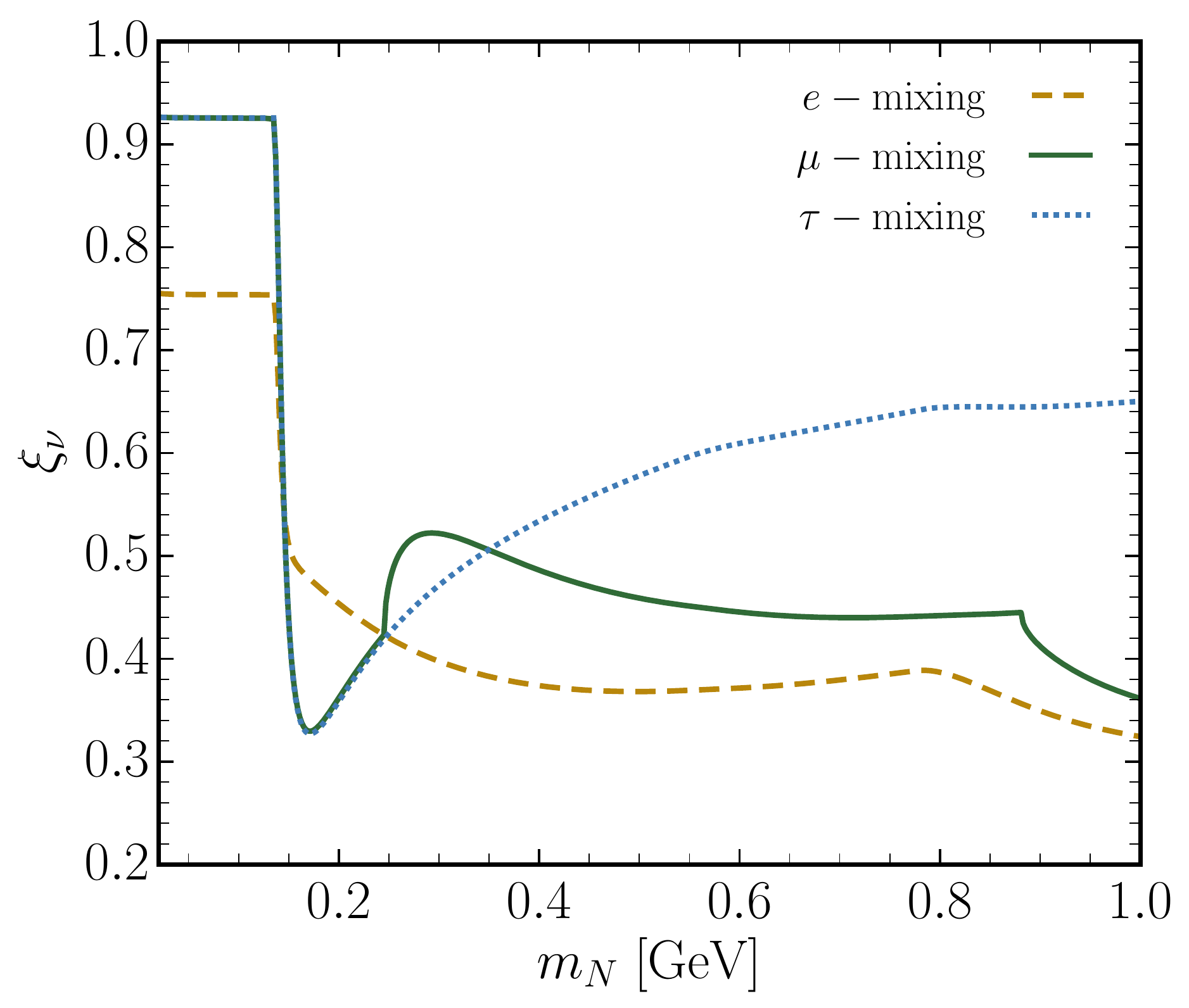}
    \caption{The fraction of HNL mass that is injected into the neutrino plasma. Contributions to this fraction from unstable HNL decay products (mesons and muons) are included and we assume that the kinetic energy of all created charged particles goes into the EM plasma. For $m_N \gtrsim 135\,\mathrm{MeV}$, HNLs can decay into neutral pions, which in their turn decay into two photons. This causes the sudden decrease of $\xi_\nu$ around that mass. At higher masses, $\xi_\nu$ keeps increasing in the case of $\tau$-mixing, which is due to the absence of HNL decays into charged mesons (such decays are possible in the other two mixing cases).}
    \label{fig:energy-injection-fraction}
\end{figure}

HNLs alter the cosmological history through their contribution to the total energy density of the Universe and their decay into SM particles. HNLs that decay well before the decoupling of active neutrinos will leave no traceable impact. On the other hand, if HNLs live long enough, they could alter several physical quantities, such as \neff and the primordial abundances of light elements~\cite{Dolgov:2000jw, Dolgov:2000pj, Dolgov:2003sg, Fuller:2011qy, Ruchayskiy:2012si, Vincent:2014rja}. Indeed, strong limits have been set on their mass and lifetime by considering their impact on Big Bang Nucleosynthesis and the Cosmic Microwave Background, see e.g.~\cite{Gelmini:2020ekg, Sabti:2020yrt, Boyarsky:2020dzc} for recent works on this subject. HNLs inject (eventually) all of their energy either into the neutrino or electromagnetic plasma. The fraction of the HNL energy that is injected into each of these two sectors is mass-dependent and shows a significant shift to the EM plasma once HNLs can decay into neutral pions (${\sim}135\,\mathrm{MeV}$), see Fig.~\ref{fig:energy-injection-fraction}.
\begin{figure}[t!]
    \vspace{-2pt}
    \centering
    \includegraphics[width=\linewidth]{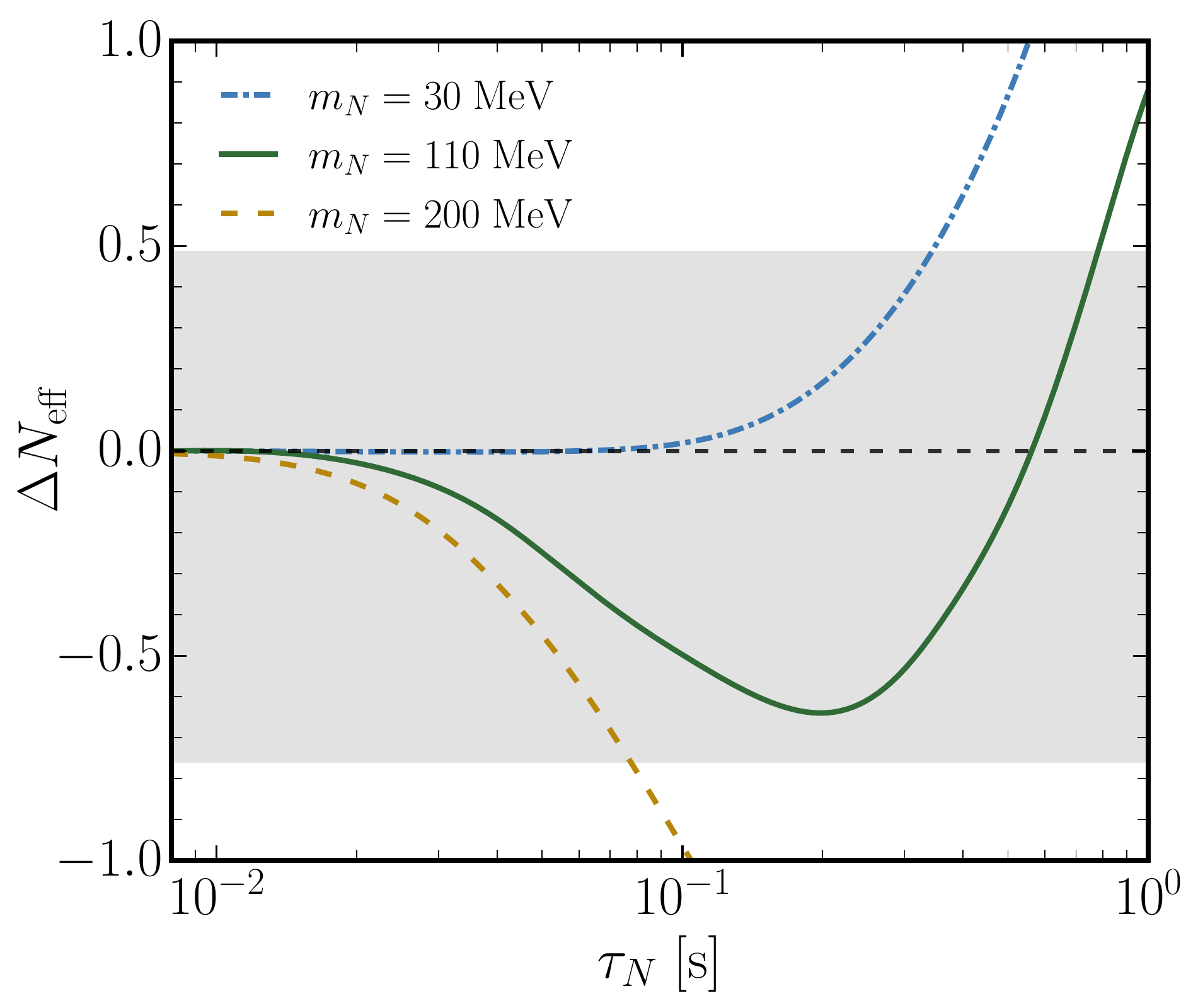}
    \caption{$\Delta\neff$ as a function of HNL lifetime for a number of benchmark masses. Mixing with electron neutrinos only is considered here. The curves illustrate three cases of how HNLs can affect \neff: \emph{1)} they can decay mostly into neutrinos and simply increase \neff ($30\,\mathrm{MeV}$ curve), \emph{2)} they can decay mostly into neutrinos and either decrease or increase \neff depending on their lifetime ($110\,\mathrm{MeV}$ curve), and \emph{3)} they can decay mostly into EM particles and simply decrease \neff ($200\,\mathrm{MeV}$ curve). HNLs with masses $m_N \gtrsim 70\,\mathrm{MeV}$ that decay mainly into neutrinos around neutrino decoupling, show an initial decrease of $\Delta\neff$ as a result of the thermalisation of the injected high-energy neutrinos. The grey band is the current sensitivity by Planck.}
    \label{fig:HNL_deltaNeff_tauN}
    \vspace{-10pt}
\end{figure}
This plot shows that HNLs below the pion mass decay mainly into neutrinos and, therefore, one would naively expect that in this mass range \neff increases. However, we find that HNLs are able to decrease \neff for masses already above ${\sim}70\,\mathrm{MeV}$, while for smaller masses an increase of \neff is observed. The origin of this sign change in $\Delta\neff$ at $m_N \gtrsim 70\,\mathrm{MeV}$ (rather than $m_N > m_\pi$ as one would guess from Fig.~\ref{fig:energy-injection-fraction}) lies in the energy transfer from the neutrino plasma to the electromagnetic plasma that is induced by the injected non-equilibrium neutrinos, as discussed earlier in Sec.~\ref{sec:qualitative-analysis}. We run \texttt{pyBBN} simulations to examine in which region of parameter space this sign change happens\footnote{We note that \texttt{pyBBN} predicts a SM value for \neff of 3.026, rather than $3.044$. This is because the code does not include higher-order QED corrections that account for a $\Delta\neff = 0.01$ increase~\cite{Bennett:2019ewm, Escudero:2020dfa, Bennett:2020zkv, Froustey:2020mcq, Akita:2020szl}, while the remaining is due to numerical inaccuracy. This, however, is only a minor difference and does not change any of the results presented in this work.}. We show $\Delta\neff$ as a function of the HNL lifetime in Fig.~\ref{fig:HNL_deltaNeff_tauN} for a number of benchmark masses. The grey band in this figure indicates the current sensitivity by Planck. Included in this figure is an HNL of mass $110\,\mathrm{MeV}$, which decreases \neff for lifetimes below $\tau_N\lesssim 0.6\,\mathrm{s}$ and increases \neff for longer lifetimes. Such a lifetime ($\tau_N \sim 0.6\,\mathrm{s}$) roughly corresponds to the time of neutrino decoupling, beyond which thermalisation between the neutrino and EM plasma is not efficient anymore and the injected neutrinos remain in the neutrino sector. This exemplifies the ability of HNLs below the pion mass to diminish \neff, even when neutrinos are on the verge of being completely decoupled. With the current sensitivity of Planck, however, this initial decrease of $\Delta\neff$ for this mass falls within the error range and is thus not observable. Nevertheless, a number of upcoming and proposed CMB missions, such as the Simons Observatory~\cite{Ade:2018sbj} and CMB-S4~\cite{Abazajian:2019eic}, could provide a determination of \neff around the percent-level and probe this effect.

\vspace{10pt}

\begin{figure}[ht!]
    \vspace{2pt}
    \centering
    \includegraphics[width=\linewidth]{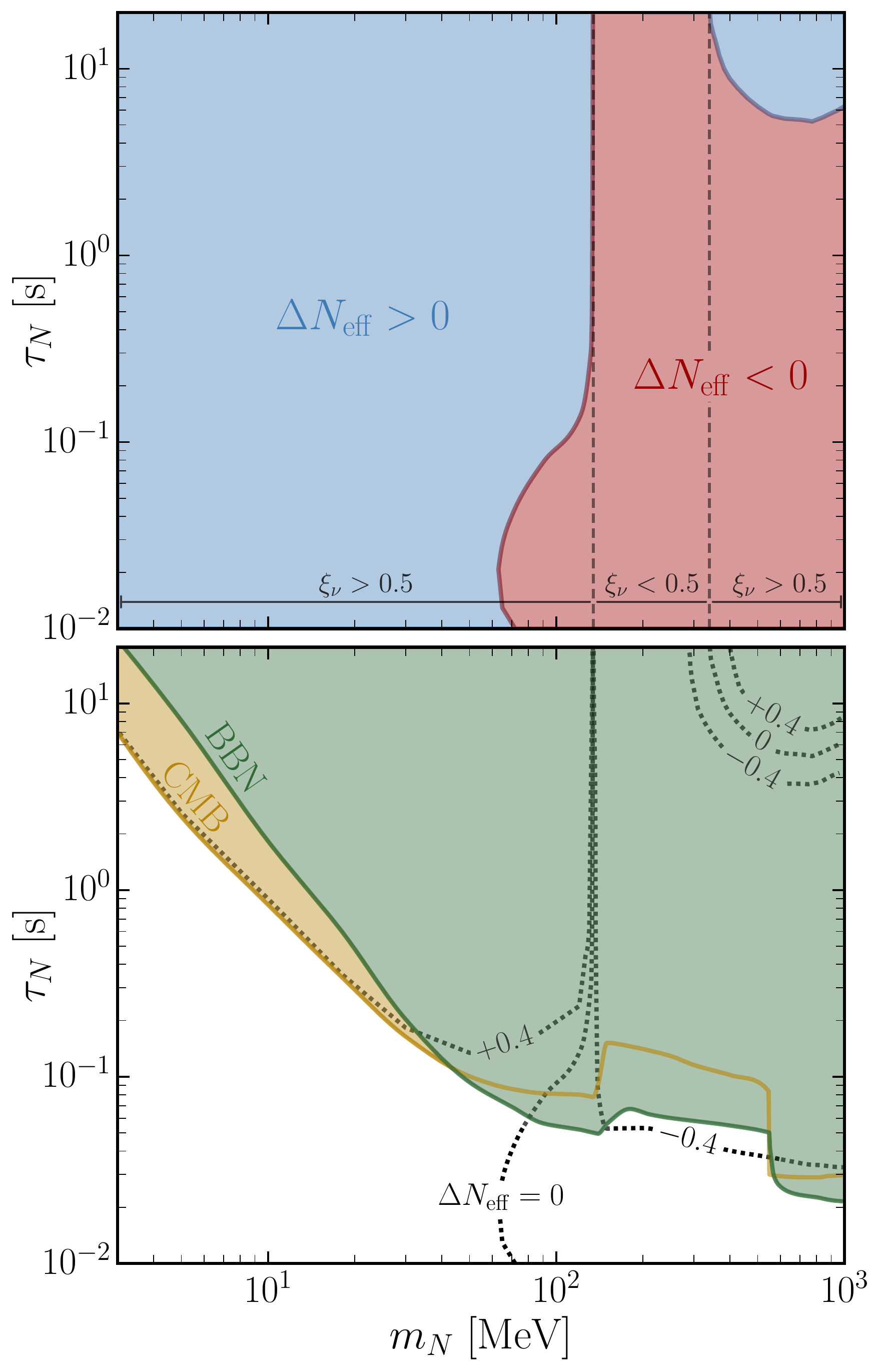}
    \caption{How HNLs change $\Delta\neff$ as a function of their mass and lifetime. Mixing with tau neutrinos only is considered here. \emph{Top panel}: Regions of the HNL parameter space that predict an increase (blue) or decrease (red) of \neff with respect to the SM value. The horizontal lines at the bottom of the plot indicate the mass ranges where HNLs inject most of their energy into neutrinos ($\xi_\nu > 0.5$) or the EM plasma ($\xi_\nu < 0.5$). In the former case, HNLs can still decrease \neff as a result of the efficient transfer of energy from neutrinos to EM particles. \emph{Bottom panel}: Regions of the HNL parameter space that are excluded by BBN abundance measurements (green) and CMB observations (yellow). The $\Delta\neff = \{0,\pm0.4\}$ contours give an indication of by how much HNLs can change \neff at the most in the unconstrained region. The BBN bound is from~\cite{Sabti:2020yrt} and uses a central value for the primordial helium abundance of $Y_\mathrm{P} = 0.245$~\cite{Zyla:2020zbs} with an error of 4.35\% (see~\cite{Boyarsky:2020dzc} for a discussion on how this error is obtained). For masses higher than the eta-meson mass (${\sim}550\,\mathrm{MeV}$), the meson effect from~\cite{Boyarsky:2020dzc} is included in the analysis. The CMB constraint is obtained using the approach as detailed in~\cite{Sabti:2020yrt} (see Sec.~\ref{subsec:HNLs_CMB} for more details on the CMB bound). This panel also shows that there is only a relatively small unconstrained region of parameter space left that can increase \neff and where HNLs could play a role in alleviating the Hubble tension.}
    \label{fig:HNL_parameterspace}
\end{figure}

We depict the region of HNL parameter space where $\Delta\neff$ changes sign in the top panel of Fig.~\ref{fig:HNL_parameterspace}. This is shown for the case of pure mixing with tau neutrinos only, as the parameter space where HNLs mix purely with electron and muon neutrinos is excluded in the lower mass range (where $\Delta\neff$ can be positive) by BBN, the CMB and experimental searches~\cite{Sabti:2020yrt, Boyarsky:2020dzc, Bondarenko:2021cpc}. In these latter two cases, $\Delta\neff$ can only be negative in the unconstrained parameter space. This top panel shows that there is a large region of HNL parameter space, where these particles inject most of their energy into neutrinos and still decrease \neff. The behaviour of negative $\Delta\neff$ continues for short-lived HNLs with masses $m_N > 1\,\mathrm{GeV}$, since the neutrino energy increases with the HNL mass. On the other hand, for HNLs with lifetimes $\tau_N \gg 1\,\mathrm{s}$, it depends on how much energy they inject into the neutrino plasma. Indeed, such HNLs decay long after neutrino decoupling, when non-equilibrium effects are not important anymore and the injected neutrinos remain in the neutrino sector. This means that the sign of $\Delta\neff$ is simply determined by the value of $\xi_\nu$. As a result, for masses where $\xi_\nu > 0.5$ (see Fig.~\ref{fig:energy-injection-fraction}) this would mean that eventually $\Delta\neff > 0$ and vice versa (see Fig.~\ref{fig:Analytic_HNLs} for an illustration).

\subsection{Bounds from the CMB}
\label{subsec:HNLs_CMB}
The CMB anisotropies are mainly sensitive to \neff through its impact on the damping tail~\cite{Hu:1996mn,Bashinsky:2003tk,Hou:2011ec, Zyla:2020zbs}. For example, a larger number of relativistic degrees of freedom causes a stronger suppression of the power spectrum at high multipoles, as temperature anisotropies below the scale of the photon diffusion length are more damped by the increased expansion rate. This effect is, however, degenerate if the primordial helium abundance $Y_\mathrm{P}$ is also considered as a free parameter~\cite{Hou:2011ec}. $Y_\mathrm{P}$ is related to the number density of free electrons\footnote{This relation between $n_e$ and $Y_\mathrm{P}$ is obtained by imposing charge neutrality on the primordial plasma. Therefore, $Y_\mathrm{P}$ is allowed to change even if the total baryon density is fixed.}, $n_e\propto (1-Y_\mathrm{P})$, which in its turn enters in the CMB damping scale. A larger $Y_\mathrm{P}$ leads to a lower electron density, a larger electron-photon interaction rate, a larger photon diffusion length and thus a stronger damping. 

\vspace{10pt}

We extend the CMB constraint on HNLs for masses up to $1\,\mathrm{GeV}$ using the same approach as detailed in~\cite{Sabti:2019mhn,Sabti:2020yrt} and show the result in the bottom panel of Fig.~\ref{fig:HNL_parameterspace}. Also included in this panel are the contours where $\Delta\neff = \pm 0.4$, which give an indication of by how much HNLs can change $\neff$ at the most, given the current constraints imposed by BBN and the CMB. The CMB bound is only stronger than the BBN bound in the lower mass range, as this is where \neff strongly increases. HNLs with short lifetimes and masses around $\mathcal{O}(10)\,\mathrm{MeV}$ decouple while being non-relativistic and thus have a suppressed number density. They can therefore survive beyond the decoupling of SM weak reactions, without significantly affecting the primordial abundances. However, since the HNL energy density here falls off as $(\text{scale factor})^{-3}$, the HNLs could eventually dominate the total energy density of the Universe. As can be seen in Fig.~\ref{fig:energy-injection-fraction}, HNLs in this lower mass range inject most of their energy into neutrinos, which remains in the neutrino sector after neutrino decoupling. The result is then a significant increase in \neff, which can be constrained with the CMB. On the other hand, for masses higher than ${\sim}70\,\mathrm{MeV}$, \neff starts decreasing. This decrease is relatively small in magnitude, especially in the region that is not constrained by BBN, where $\neff - \neff^\mathrm{CMB} \lesssim 0.4$. In addition, the error in the determination of $Y_\mathrm{P}$ by the CMB is larger than the one by BBN~\cite{Aghanim:2018eyx, Sabti:2019mhn}. These two properties make the CMB a weaker probe of HNLs in the higher mass range. As mentioned before, future CMB experiments could improve upon this result.

\subsection{Implications for the Hubble Parameter}
\label{subsec:Hubble_param}
\vspace{-2pt}

An increase or decrease of \neff subsequently also changes the Hubble parameter. As such, HNLs could play a role in alleviating the longstanding tension between local determinations of the current day Hubble rate $H_0$ and the one as inferred from the CMB\footnote{This question has been considered before in~\cite{Gelmini:2019deq}. Importantly, this study used the results of~\cite{Dolgov:2000jw}, where the assumption was made that any change in the primordial helium abundance is due to $\Delta\neff$. In contrast, here we find that neutrino spectral distortions are the driving power behind $\Delta\neff$ for short-lived HNLs. As a consequence, the results presented in our work and in~\cite{Gelmini:2019deq} are rather different.}~\cite{Verde:2019ivm, Efstathiou:2020wxn}. The usual approach involves increasing \neff, while keeping the angular scale of the sound horizon $\theta_\mathrm{s} = r_\mathrm{s}/D_\mathrm{A}$ fixed, see e.g.~\cite{Knox:2019rjx,Vagnozzi:2019ezj, DiValentino:2021izs}. Here, $r_\mathrm{s}$ is the comoving sound horizon and $D_\mathrm{A}$ is the comoving angular diameter distance to the surface of last scattering. Both of these quantities depend on the Hubble parameter:

\begin{figure}[t!]
    \centering
    \includegraphics[width=\linewidth]{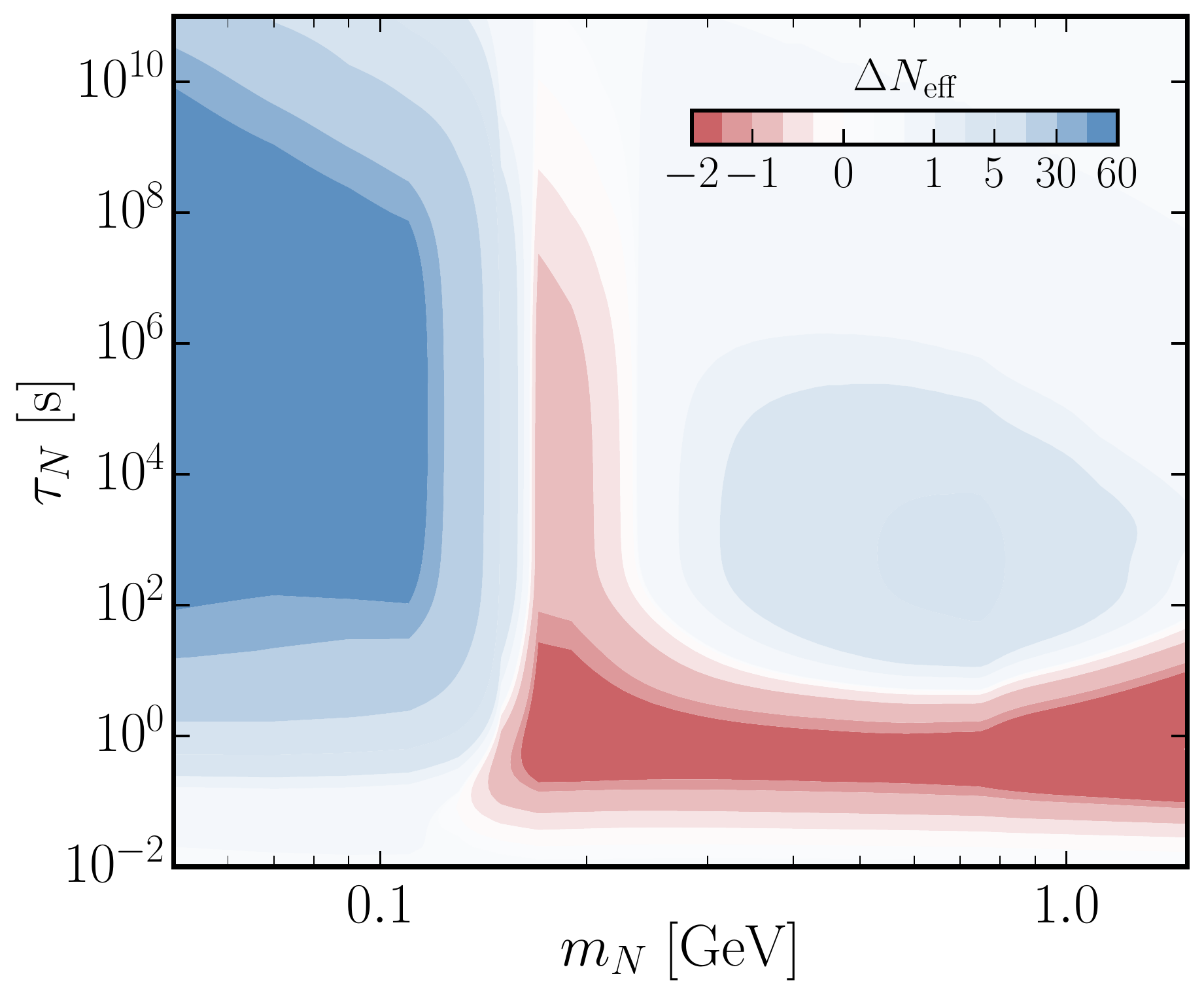}
    \caption{Semi-analytic estimate of $\Delta\neff$ as a function of HNL mass and lifetime in the case of pure tau mixing. This plot is obtained using the method described in Sec.~\ref{sec:qualitative-analysis} (and is therefore only accurate up to a factor $3-4$ for short lifetimes, when neutrinos are still in partial equilibrium). Nevertheless, it allows for a qualitative understanding of the behaviour of $\Delta\neff$ at lifetimes larger than considered in the main analysis (Fig.~\ref{fig:HNL_parameterspace}). Importantly, for lifetimes well beyond the time of neutrino decoupling ($\mathcal{O}(1)\,\mathrm{s}$), non-equilibrium effects are absent and the sign of $\Delta\neff$ is thus completely determined by the fraction of HNL energy $\xi_\nu$ that is injected into the neutrino plasma, see Fig.~\ref{fig:energy-injection-fraction}. We see that HNLs with low masses and long lifetimes can still considerably affect \neff, while in the higher mass range $\Delta\neff$ tends to 0. This is because low-mass HNLs are more abundantly produced in this region of parameter space~\cite{Gelmini:2020ekg}, where their mixing angles are relatively large.}
    \label{fig:Analytic_HNLs}
\end{figure}

\begin{align}
    \label{eq:rs}
    r_\mathrm{s}(z_*) &= \int_{z_*}^{\infty}\frac{c_s(z)\mathrm{d}z}{H(z)}\\
    \label{eq:DA}
    D_\mathrm{A}(z_*) &= \int_{0}^{z_*}\frac{\mathrm{d}z}{H(z)}\, ,
\end{align}
where $z_*$ is the redshift of the last-scattering surface and $c_\mathrm{s}(z)$ is the speed of sound of the baryon-photon fluid in the early Universe. The Hubble rate in Eq.~\eqref{eq:rs} depends mainly on the radiation (photons \emph{and} neutrinos) and matter energy densities, while the one in Eq.~\eqref{eq:DA} is the late-time Hubble rate and depends mostly on the dark energy and matter energy densities. This means that increasing \neff only results in a larger early-time Hubble rate and a smaller $r_\mathrm{s}$. In order to keep $\theta_\mathrm{s}$ fixed, the comoving angular diameter distance must satisfy $D_\mathrm{A} = r_\mathrm{s}/\theta_\mathrm{s}$, which then also decreases if $r_\mathrm{s}$ decreases. Looking at Eq.~\eqref{eq:DA}, such a decrease can be accomplished by increasing the dark energy density $\omega_\Lambda$, or equivalently, $H_0$ (as $\Omega_\Lambda = 1 - \Omega_\mathrm{m}$). Since local measurements find a larger value of $H_0$ than the one inferred from the CMB within the Standard Model, this approach provides a way to reduce the Hubble tension.

\vspace{10pt}

This method, however, does not take into account the increased Silk damping induced by a larger \neff~\cite{Hu:1996mn,Bashinsky:2003tk,Hou:2011ec, Zyla:2020zbs}. Therefore, a price must be paid when alleviating the Hubble tension in this way: An increase of \neff leads to a larger disagreement with the CMB itself. Given our CMB constraint in Fig.~\ref{fig:HNL_parameterspace}, we see that HNLs can increase \neff by at most $\Delta\neff \approx 0.4$. This gives us an indication of the extent to which unconstrained HNLs could increase $H_0$ and ameliorate the Hubble tension. We estimate the corresponding $H_0$ by running \texttt{Monte Python}~\cite{Audren:2012wb, Brinckmann:2018cvx} with the Planck 2018 baseline TTTEEE+lowE analysis. Fixing the primordial helium abundance to\footnote{This is approximately the value of $Y_\mathrm{P}$ along the $\Delta\neff = +0.4$ curve on the left in the bottom panel of Fig.~\ref{fig:HNL_parameterspace}.} $Y_\mathrm{P} = 0.25$, we obtain\footnote{All errors in $H_0$ reported here are at 68\% CL.} $H_0 = 70.5\pm 0.7\,\mathrm{km\,s^{-1}Mpc^{-1}}$. This value can be compared to the one as obtained from, e.g., a distance ladder approach, which gives $H_0^\mathrm{local} = 73.0\pm 1.4\,\mathrm{km\,s^{-1}Mpc^{-1}}$~\cite{Riess:2020fzl}. Given the Hubble rate obtained within $\Lambda$CDM ($H_0 = 67.3\pm 0.6\,\mathrm{km\,s^{-1}Mpc^{-1}}$~\cite{Aghanim:2018eyx}), we see that HNLs which are not excluded by BBN, the CMB and terrestrial experiments can moderately alleviate the Hubble tension.

\section{Conclusions}
\label{sec:conclusions}
In this work, we have studied how heavy, unstable FIMPs that can decay into neutrinos impact the number of relativistic species \neff in the early Universe. A particularly interesting effect that could occur with these particles, is when they inject most of their energy into neutrinos but still decrease \neff. This could happen if FIMPs decay when neutrinos are still in (partial) equilibrium ($\tau_\mathrm{FIMP} \sim \mathcal{O}(0.1)\,\mathrm{s}$) and is a direct consequence of the thermalisation process of the injected high-energy neutrinos (see Sec.~\ref{sec:qualitative-analysis} for a semi-analytical treatment of this effect). Here we identify neutrino spectral distortions as the driving power behind this effect, since they lead to an efficient transfer of energy from the neutrino plasma to the electromagnetic plasma (see Figs.~\ref{fig:Neff_analytic} and~~\ref{fig:rhonu_rhoEM_inst}). Some of the injected neutrino energy gets quickly transferred to the EM plasma, while the remaining will stay as residual spectral distortions in the neutrino distribution functions. These spectral distortions keep the energy transfer balance of $\nu\leftrightarrow\mathrm{EM}$ reactions shifted to the right till long after FIMP decay. In order to accurately account for this effect, it is therefore important to solve the Boltzmann equation and track the evolution of the neutrino distribution functions. Using a thermal-like distribution for neutrinos as an approximation can lead to incorrect results, e.g., that \neff can never decrease when FIMPs inject most of their energy into neutrinos.

\vspace{10pt}

From our simulations, done with the publicly available Boltzmann code \texttt{pyBBN}~\cite{Sabti:2020yrt}, we find that this mechanism is especially relevant for FIMPs that can decay into neutrinos with average energies $E_\nu^\mathrm{inj} \gtrsim  25\,\mathrm{MeV}$. In case such neutrinos are created via 2- or 3-body decays, this roughly corresponds to FIMP masses $m_\mathrm{FIMP}^\text{2-body} \gtrsim 50\,\mathrm{MeV}$ and $m_\mathrm{FIMP}^\text{3-body} \gtrsim 70\,\mathrm{MeV}$ respectively. This is in agreement with the results presented in~\cite{Hannestad:2004px}. As such, this effect may be relevant for many classes of FIMPs\footnote{While \texttt{pyBBN} is mainly built to simulate the cosmological history in the presence of Heavy Neutral Leptons, it can in principle be modified to include many other classes of FIMPs.}, including Higgs-like dark scalars~\cite{Boiarska:2019jym}, dark photons~\cite{Fabbrichesi:2020wbt}, neutralinos in supersymmetric models with broken R-parity~\cite{Dumitru:2018jyb}, vector portals coupled to anomaly-free currents~\cite{Kling:2020iar} and short-lived neutrinophilic scalars~\cite{Kelly:2019wow}.

\vspace{10pt}

As a case study, we have considered Heavy Neutral Leptons and illustrated for which masses and lifetimes the aforementioned effect occurs. We show this in the top panel of Fig.~\ref{fig:HNL_parameterspace} for the case of pure mixing with tau neutrinos. Such particles can decrease \neff for masses already above ${\sim}70\,\mathrm{MeV}$, even if they inject most of their energy into neutrinos. HNLs that mix purely with electron or muon neutrinos can only decrease \neff, as the parameter space in which they increase \neff is already excluded by BBN, CMB and experimental constraints~\cite{Sabti:2020yrt, Boyarsky:2020dzc}. Therefore, in the pure mixing cases, only HNLs that mix with tau neutrinos have an unconstrained region of parameter space left where $\Delta\neff$ can be positive (bottom panel of Fig.~\ref{fig:HNL_parameterspace}). Such HNLs can increase \neff by at most $\Delta\neff \approx +0.4$, as higher values are excluded by BBN and the CMB. Given this maximum allowed value of $\Delta\neff$, we then estimated what current-day Hubble rate it would correspond to. Using the Planck 2018 baseline TTTEEE+lowE analysis, we find that HNLs can increase the Hubble rate to at most $H_0 = 70.5\pm 0.7\,\mathrm{km\,s^{-1}Mpc^{-1}}$ and therefore moderately alleviate the Hubble tension.


\section*{Acknowledgements}
We thank Oleg Ruchayskiy for his collaboration at the beginning of this project. We are grateful to James Alvey and Miguel Escudero for insightful comments on the draft of this work, and to Steen Hannestad for valuable discussions on the thermalisation of neutrinos. AB, MO and VS have received funding from the European Research Council (ERC) under the European Union’s Horizon 2020 research and innovation program (GA 694896). VS has also received funding from the Carlsberg foundation. NS is a recipient of a King's College London NMS Faculty Studentship. We acknowledge the use of the public cosmological codes \texttt{CLASS}~\cite{Blas:2011rf,Lesgourgues:2011re} and \texttt{Monte Python}~\cite{Audren:2012wb, Brinckmann:2018cvx}.


\bibliographystyle{apsrev4-1}
\bibliography{biblio}

\appendix

\section{Effect of Residual Non-equilibrium Neutrino Distortions}
\label{app:residual_spectral}
The simple model described in Sec.~\ref{sec:qualitative-analysis} relies on the assumption that the remaining fraction $1-\epsilon$ of the injected neutrino energy is perfectly thermal. In reality, this may not be the case and the full thermalisation would occur during a much larger number of interactions than $N_{\text{therm}} \simeq \log_{2}(E_\nu^\text{inj}/3.15T)$. Therefore, this simple model underestimates the energy fraction that goes into the EM plasma\footnote{Once the energy of the non-equilibrium neutrinos is close to the average thermal energy of $3.15T$, they lose roughly $\Delta E_{\nu} = (E_{\nu}-3.15T)/2$ of energy per scattering. Therefore, the number of scatterings required to diminish $E_{\nu}$ down to $3.15T$ is larger.}. The remaining non-equilibrium neutrinos will manifest themselves as residual non-thermal spectral distortions in the distribution function of neutrinos. These spectral distortions keep the energy exchange balance of $\nu\leftrightarrow \EM$ reactions shifted to the right till long after FIMP decay. As a result, more neutrino energy will be transferred to the EM plasma and \neff can further decrease.
There is a subtlety here that the remaining $1-\epsilon$ non-equilibrium neutrinos are only slightly hotter than the thermal neutrinos, and we cannot describe their thermalisation as an instant process: The corresponding rate is comparable to the thermal energy exchange rate. As such, the energy transfer process is extended in time, and a proper study of this effect requires solving the Boltzmann equation for the neutrino distribution function. 

\begin{figure*}
    \centering
    \includegraphics[width=0.93\linewidth]{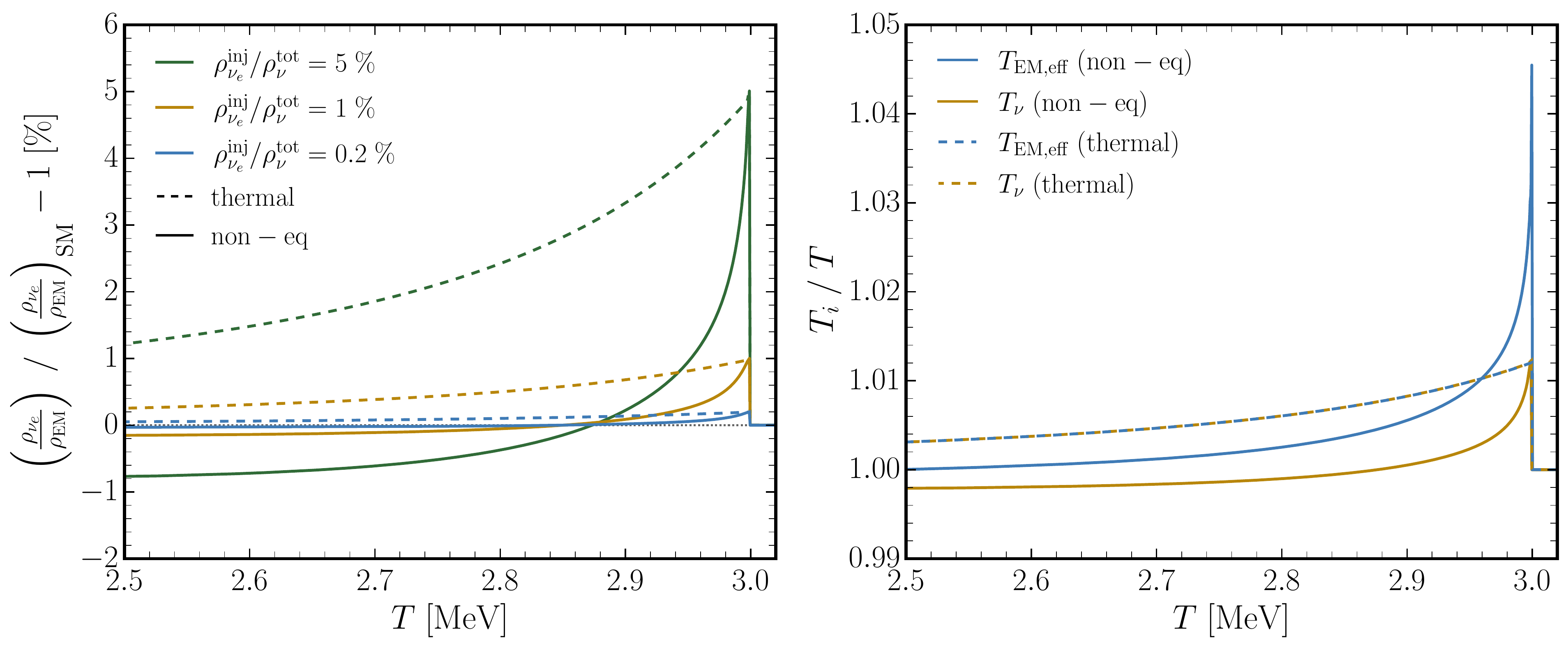}
    \caption{Evolution of the neutrino and EM plasma after the instant injection of neutrinos with energy $E_\nu^\text{inj} = 70\,\mathrm{MeV}$ at $T = 3\,\mathrm{MeV}$. \emph{Left panel}: The ratio of electron neutrino energy density to electromagnetic energy density, relative to the SM prediction. Three fractions of the injected energy density are considered: $\rho_{\nu_e}^\text{inj}/\rho_{\nu}^\mathrm{tot} = \{0.2\%
    ,1\%,5\%\}$. The solid lines are obtained by taking into account the full non-equilibrium spectrum of neutrinos, whereas the dashed lines correspond to the evolution assuming that neutrinos always have a thermal-like spectrum with temperature $T_{\nu}  \propto \rho_{\nu}^{1/4}$. \emph{Right panel}: Evolution of the neutrino temperature (dashed) and effective EM plasma temperature (solid) for which the energy transfer rate in Eq.~\eqref{eq:energy-transfer-rate} vanishes. An injected fraction of $\rho_{\nu_e}^\text{inj}/\rho_{\nu}^\mathrm{tot} = 5\%$ is considered here. The solid and dashed lines indicate when non-equilibrium and thermal-like neutrino distributions are used respectively.}
    \label{fig:rhonu_rhoEM_inst}
\end{figure*}

\vspace{10pt}

To study the impact of neutrino spectral distortions on the $\nu\to\mathrm{EM}$ energy balance shift, we consider a simple scenario where high-energy neutrinos are instantly injected into the primordial plasma. We make use of the publicly available Boltzmann code \texttt{pyBBN}\footnote{\href{https://github.com/ckald/pyBBN}{https://github.com/ckald/pyBBN}}~\cite{Sabti:2020yrt} to simulate this process and to track the evolution of the neutrino distribution functions. Within this setup, neutrinos with energy $E_\nu^\mathrm{inj} = 70\,\mathrm{MeV}$ are instantly injected at $T=3\,\mathrm{MeV}$. They amount for a fixed percentage of the total neutrino energy density and are equally distributed over the three neutrino flavours. All Standard Model interactions as specified in~\cite{Sabti:2020yrt} are included, but with neutrino oscillations turned off (without any loss of generality). In order to highlight the importance of neutrino spectral distortions, we perform this procedure a second time, but with neutrino spectral distortions turned off. In that case, the neutrino distribution function is given by a Fermi-Dirac distribution with temperature $T_{\nu_\alpha} = \left(\frac{240\rho_{\nu_\alpha}}{7\pi^2g_{\nu_\alpha}}\right)^{1/4}$, where $\rho_{\nu_\alpha}$ and $g_{\nu_\alpha} = 2$ are the energy density (of both neutrinos and anti-neutrinos) and number of degrees of freedom of neutrino flavour $\alpha$ respectively.

\vspace{10pt}

The evolution of the ratio $\rho_{\nu_e}/\rho_{\text{EM}}$ (relative to the one in the SM) is shown in the left panel of Fig.~\ref{fig:rhonu_rhoEM_inst} for different amounts of injected neutrino energy. In agreement with the story in Sec.~\ref{sec:qualitative-analysis}, we observe a fast drop-off in the ratio right after the injection, which signifies the quick transfer of energy from the neutrino plasma to the EM plasma. After reaching the SM value (which naively corresponds to an equilibrium state), the ratio continues decreasing. This is the effect of the extended thermalisation due to neutrino spectral distortions, as caused by the remaining fraction $1-\epsilon$ of non-equilibrium neutrinos. Eventually, the ratio will be smaller than the SM value and $\Delta\neff$ becomes negative. In this plot, the dashed lines correspond to the same simulations but with a thermal-like distribution for the neutrinos. It is clear that without spectral distortions, the energy transfer from the neutrino sector to the EM sector is much less efficient.

\vspace{10pt}

Another way to look at this shift in the energy transfer balance from the neutrino plasma to the EM plasma is to ask the question: Which temperature $T_\mathrm{EM,eff}$ is the EM plasma \emph{trying} to reach after the injection? As we will see, depending on whether neutrinos have a non-equilibrium or a thermal-like distribution, this temperature can be either larger than or equal to the neutrino temperature\footnote{In all cases, with `neutrino temperature' we refer to the quantity $T_\nu = \left(\frac{240\rho_\nu}{7\pi^2g_\nu}\right)^{1/4}$, where $g_\nu = 2$ and $\rho_\nu$ is the energy density of both neutrinos and anti-neutrinos.}. In the former case, it means that the EM plasma temperature can exceed the neutrino temperature (and thus $\Delta\neff$ can be negative), while in the latter case $\Delta\neff$ \emph{cannot} be negative.

\vspace{10pt}

In more technical terms, the exchange of energy between neutrinos and EM particles is regulated by the Boltzmann collision integral $I_\mathrm{coll}$, which encodes all interactions between the species. For neutrinos that participate in reactions of the form $\nu + 2 \leftrightarrow 3 + 4$, the collision integral is given by~\cite{Kolb:1990vq}:
\begin{align}
\label{eq:CollisionIntegral}
I_\nu = & \frac{1}{2g_\nu E_\nu}\sum_{\mathrm{reactions}}\int\prod_{i=2}^{4}\left(\frac{\mathrm{d}^3p_i}{(2\pi)^32E_i}\right)|\mathcal{M}|^2\times \nonumber\\
& \times \left[(1-f_\nu)(1-f_2)f_3f_4 - f_\nu f_2(1-f_3)(1-f_4) \right]\times \nonumber\\
& \times(2\pi)^4\delta^4(P_\nu + P_2 - P_3 - P_4)\ ,
\end{align}

where $f_i$ and $P_i$ are the distribution function and four momentum of species $i$ respectively, and $|\mathcal{M}|^2$ is the unaveraged squared matrix element summed over degrees of freedom of initial and final states. The energy transfer rate between the neutrino and EM plasma can be written as:
\begin{align}
    \Gamma(T_{\text{EM}}) = \int \frac{\mathrm{d}^{3}p_{\nu}}{(2\pi)^{3}} I_{\text{coll}}(T_\mathrm{EM})E_{\nu}\ ,
    \label{eq:energy-transfer-rate}
\end{align}

where we consider $I_{\text{coll}}$ to be a function of the EM plasma temperature $T_\mathrm{EM}$. There exists a temperature $T_{\EM,\eff}$ for which this rate is equal to 0. This corresponds to the temperature the EM plasma \emph{tends} to during thermalisation, since then the system would be in equilibrium. In the case where neutrinos would have a thermal-like spectrum with temperature $T_{\nu}$, the rate vanishes when $T_{\EM,\eff} = T_{\nu}$. On the other hand, when a non-equilibrium neutrino spectrum is considered, we find that $T_{\EM,\eff} > T_{\nu}$ when $\Gamma = 0$. In the former case \neff \emph{cannot} decrease, while in the latter case the EM plasma temperature can exceed $T_\nu$ and \neff can thus decrease. We show the evolution of $T_\mathrm{EM,eff}$ and $T_\nu$ as obtained from the instant neutrino injection simulations in the right panel of Fig.~\ref{fig:rhonu_rhoEM_inst}. 

\vspace{10pt}

The conclusion here is that neutrino spectral distortions play a central role in transferring energy from the neutrino sector to the EM sector. When considering short-lived FIMPs that can decay into neutrinos, the impact of these distortions on the evolution of \neff should not be neglected.

\section{Temperature Evolution Equations}
\label{app:T_eqs}
Here we provide the relevant equations for the time evolution of the neutrino and photon temperatures in the presence of decaying FIMPs. Assuming a Fermi-Dirac distribution for neutrinos, the equations read~\cite{Escudero:2018mvt, Escudero:2020dfa}:

\newpage

\begin{align}
    & \frac{dT_{\nu}}{dt} + 4HT_{\nu} = \frac{(1-\xi_{\EM,\eff})\frac{\rho_{\text{FIMP}}}{\tau_{\text{FIMP}}}+\Gamma_{\nu\leftrightarrow \EM}(T_{\nu},T_{\EM})}{d\rho_{\nu}/dT_{\nu}} \\ 
    &\frac{dT_{\EM}}{dt} + \frac{(4H\rho_{\gamma}+3H(\rho_{e}+p_{e}))}{d\rho_{e}/dT + d\rho_{\gamma}/dT} \nonumber \\
    &\hspace{2cm} = \frac{\xi_{\EM,\eff}\frac{\rho_{\text{FIMP}}}{\tau_{\text{FIMP}}}-\Gamma_{\nu\leftrightarrow \EM}(T_{\nu},T_{\EM})}{d\rho_{e}/dT + d\rho_{\gamma}/dT} \\ 
    &\frac{d\rho_{\text{FIMP}}}{dt}+3H\rho_{\text{FIMP}} =-\frac{\rho_{\text{FIMP}}}{\tau_{\text{FIMP}}}\ ,  
\end{align}
where $\xi_\mathrm{EM,eff}$ is given in Eq.~\eqref{eq:fEM_eff}, $\rho_i$ is the energy density of particle $i$, $\tau_\mathrm{FIMP}$ is the FIMP lifetime and $\Gamma_{\nu\leftrightarrow \EM}(T_{\nu},T_{\EM}) = \left(\Gamma_{\nu_{e}\leftrightarrow \EM}+2\Gamma_{\nu_{\mu}\leftrightarrow \EM}\right)/3$ is the energy density exchange rate averaged over neutrino flavours, given by Eqs.~(2.12a) and~(2.12b) in~\cite{Escudero:2018mvt}.

\section{Fitting Functions for $N_\mathrm{eff}$}
\label{app:HNL_fitting_functions}
We summarise \texttt{pyBBN} predictions for $N_\mathrm{eff}$ in the form of fitting functions for the three pure HNL mixing cases. This may provide a quick way to predict the impact of HNLs on several cosmological probes through the change in $N_\mathrm{eff}$. They read:

\newpage

\begin{align}
    \Delta N_\mathrm{eff}^\mathrm{Fit}\Big|_\mathrm{e-mixing} &= - \frac{9.78 \tau_N e^{5.72\tau_N}}{1 + \frac{1.28\cdot10^5}{m_N^{2.42}}}\\
    \Delta N_\mathrm{eff}^\mathrm{Fit}\Big|_\mathrm{\mu-mixing} &= - \frac{7.49 \tau_N e^{12.1\tau_N}}{1 + \frac{2.41\cdot10^6}{m_N^{2.87}}}\\
    \Delta N_\mathrm{eff}^\mathrm{Fit}\Big|_\mathrm{\tau-mixing} &= - \frac{8.72 \tau_N e^{13.9\tau_N}}{1 + \frac{3.49\cdot10^3}{m_N^{1.51}}}\ ,
\end{align}

where $m_N$ is the HNL mass in MeV and $\tau_N$ is the HNL lifetime in seconds. The change in \neff is with respect to the SM value of $N_\mathrm{eff}^\mathrm{SM} = 3.026$. The fitting functions are tested for masses $100\,\mathrm{MeV} \leq m_N \leq 1\,\mathrm{GeV}$ and lifetimes $0.02\,\mathrm{s} \leq \tau_N \leq 0.05\,\mathrm{s}$, and have a maximum deviation from the simulated data of roughly ${\sim}3\%$.

\section{Comment on ``Massive sterile neutrinos in the early universe: From thermal decoupling to cosmological constraints'' by Mastrototaro et al.}
\label{app:previous_literature}
After our work was submitted, the paper~\cite{Mastrototaro:2021wzl} appeared that studies the impact of HNLs with masses $m_N < m_\pi$ on \neff. The authors of this work used numerical simulations in order to obtain \neff and disagree with our conclusion that \neff can decrease even if most of the HNL energy is injected into neutrinos. They have presented an analytic argument in their Appendix~C which aims to demonstrate that our conclusion on \neff is wrong. They start with a toy model in Eq.~(C.1) that describes the evolution of the distribution function of neutrinos $f_{\nu}$:
\begin{equation}
    x\partial_{x}f_{\nu}(E_{\nu},x) = \frac{1}{H}\left[S(x,E_{\nu}) + \varsigma^2 G_{F}^{2}T^{4}E_{\nu}( f_{\text{eq}}-f_{\nu})\right]\ ,
    \label{eq:ref-equation}
\end{equation}
where $x = ma$ (with $a$ the scale factor and $m = 1\,\mathrm{MeV}$), $H$ is the Hubble rate, $\varsigma$ is a constant and $S(x)>0$ is the source term from decays of HNLs. The second term in the brackets describes the interactions between neutrinos and EM particles, where $f_{\text{eq}}$ is the equilibrium distribution function resulting from the interaction dynamics of neutrinos and EM particles in the presence of HNLs.{\parfillskip=0pt\par}

\newpage

Their argument as to why \neff cannot decrease goes as follows: as far as the source injecting rate $S(x,E_\nu)$ and the collision rate $G_{F}^{2}T^{4}E_{\nu}$ are much higher than the Hubble rate, the solution of Eq.~\eqref{eq:ref-equation} may be given in terms of the quasi-static solution:
\begin{align}
    f_{\nu} \approx f_{\text{eq}}+\frac{S}{G_{F}^{2}T^{4}E_{\nu}}\ .
\end{align}
In the limiting case $S \ll G_{F}^{2}T^{4}E_{\nu}$, the solution is just $f_{\nu} = f_{\text{eq}}$, while in the opposite case $f_{\nu} \gg f_{\text{eq}}$. The authors conclude that in any case $f_{\nu}\geq f_{\text{eq}}$ and thus $\Delta \neff \geq 0$. However, while this argument may be applicable at very early times when neutrinos are in perfect equilibrium, it is no longer valid at temperatures $T = \mathcal{O}(1\text{ MeV})$, when they start to decouple. During the decoupling process, the dynamics of the equilibration between neutrinos and EM particles, i.e., the energy transfer between the two sectors, becomes very important and is not captured by Eq.~\eqref{eq:ref-equation}.

\vspace{10pt}

We reiterate our argument as to why \neff can decrease when FIMPs inject most of their energy into neutrinos, but now from the point of view of the neutrino distribution function (see also the right panel of Fig.~\ref{fig:rhonu_rhoEM_inst} and the surrounding text for a similar discussion). Before the decay of the FIMP, the neutrino distribution function is the same as the equilibrium distribution, $f_\nu = f_\mathrm{eq}$. Right after the decay of the FIMP, the neutrino distribution at high energies becomes $f_\nu > f_\mathrm{eq}$, while at low energies it is still $f_\nu = f_\mathrm{eq}$. During the thermalisation, high-energy neutrinos interact with both low-energy neutrinos and EM particles. In this process, the temperature of the equilibrium distribution function $f_\mathrm{eq}$ increases. Now, neutrinos in the high-energy tail of $f_\nu$ interact efficiently, see Eq.~\eqref{eq:noneq_eq_rates}, and $f_\nu \longrightarrow f_\mathrm{eq}$ for such neutrinos. But at low energies, neutrinos do not interact efficiently anymore to catch up with the increase of $f_\mathrm{eq}$, which eventually leads to $f_\nu < f_\mathrm{eq}$ in this energy range. Given that these low-energy neutrinos contribute the most to \neff, it means that $\Delta \neff$ can become negative.

\end{document}